\documentclass[sigconf]{acmart}
\AtBeginDocument{%
  }

\def\markup{0}

\if\markup1
\newcommand{\rv}[1]{{\leavevmode\color{blue}#1}}
\else
\newcommand{\rv}[1]{#1}
\newcommand{\st}[1]{}
\newcommand{\sout}[1]{}
\fi

\copyrightyear{2025}
\acmYear{2025}
\setcopyright{acmlicensed}\acmConference[CHI '25]{CHI Conference on Human Factors in Computing Systems}{April 26-May 1, 2025}{Yokohama, Japan}
\acmBooktitle{CHI Conference on Human Factors in Computing Systems (CHI '25), April 26-May 1, 2025, Yokohama, Japan}
\acmDOI{10.1145/3706598.3714236}
\acmISBN{979-8-4007-1394-1/25/04}

\usepackage{multirow} 
\usepackage{xcolor}

\definecolor{blue}{HTML}{2D9CDB}
\definecolor{green}{HTML}{06BA49}
\definecolor{yellow}{HTML}{F79600}

\begin{document}

\title[RemoteChess: A VR Chinese Chess Community for Elder Connectedness]{RemoteChess: Enhancing Older Adults' Social Connectedness via Designing a Virtual Reality Chinese Chess (Xiangqi) Community
}

\author{Qianjie Wei}
\affiliation{%
  \institution{The Hong Kong University of Science and Technology (Guangzhou)}
  \city{Guangzhou}
  \country{China}}
\email{qwei883@connect.hkust-gz.edu.cn}
\orcid{0009-0001-4429-4499}

\author{Xiaoying Wei}
\affiliation{%
  \institution{The Hong Kong University of Science and Technology}
  \city{Hong Kong SAR}
  \country{China}}
\email{xweias@connect.ust.hk}
\orcid{0000-0003-3837-2638}

\author{Yiqi Liang}
\affiliation{%
  \institution{The Hong Kong University of Science and Technology (Guangzhou)}
  \city{Guangzhou}
  \country{China}}
\email{yliang339@connect.hkust-gz.edu.cn}
\orcid{0009-0005-2596-5428}

\author{Fan Lin}
\affiliation{%
  \institution{The Hong Kong University of Science and Technology (Guangzhou)}
  \city{Guangzhou}
  \country{China}}
\email{flin040@connect.hkust-gz.edu.cn}
\orcid{0009-0000-5965-4840}

\author{Nuonan Si}
\affiliation{%
  \institution{The Hong Kong University of Science and Technology (Guangzhou)}
  \city{Guangzhou}
  \country{China}}
\email{nsi226@connect.hkust-gz.edu.cn}
\orcid{0009-0008-8478-7125}

\author{Mingming Fan}
\authornote{Corresponding author}
\affiliation{%
  \institution{The Hong Kong University of Science and Technology (Guangzhou)}
  \city{Guangzhou}
  \country{China}}
\affiliation{%
  \institution{The Hong Kong University of Science and Technology}
  \city{Hong Kong SAR}
  \country{China}}
\email{mingmingfan@ust.hk}
\orcid{0000-0002-0356-4712}

\renewcommand{\shortauthors}{Wei, et al.}

\begin{abstract}
The decline of social connectedness caused by distance and physical limitations severely affects older adults' well-being and mental health. While virtual reality (VR) is promising for older adults to socialize remotely, existing social VR designs primarily focus on verbal communication (e.g., reminiscent, chat). Actively engaging in shared activities is also an important aspect of social connection. We designed RemoteChess, which constructs a social community and a culturally relevant activity (i.e., Chinese chess) for older adults to play while engaging in social interaction. We conducted a user study with groups of older adults interacting with each other through RemoteChess. Our findings indicate that RemoteChess enhanced participants’ social connectedness by offering familiar environments, culturally relevant social catalysts, and asymmetric interactions. We further discussed design guidelines for designing culturally relevant social activities in VR to promote social connectedness for older adults.
\end{abstract}



\begin{teaserfigure}
  \includegraphics[width=\textwidth]{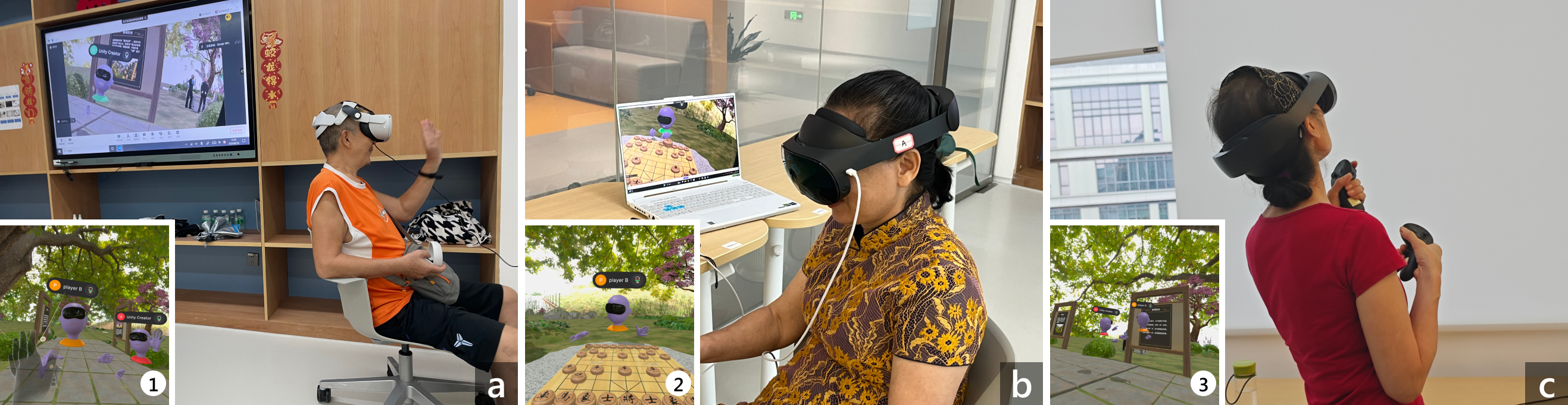}
  \caption{The figure shows three older adults are experiencing the RemoteChess VR community remotely. (a) Waving to say hello to others. (b) Playing Chinese chess with remote partners. (c) Visiting cultural exhibitions related to Chinese chess. Images in the bottom left corner show user's view in VR (1) (2) (3).}
  \Description{The figure shows three older adults are experiencing the RemoteChess VR community remotely. (a) Waving to say hello to others. (b) Playing Chinese chess with remote partners. (c) Visiting cultural exhibitions related to Chinese chess. Images in the bottom left corner show user's view in VR (1) (2) (3).}
  \label{fig:teaser}
\end{teaserfigure}

\maketitle

\section{Introduction}

Social connectedness plays a vital role in improving older adults' mental and physical health, effectively reducing feelings of loneliness and maintaining cognitive abilities \cite{ashida2008differential, holt2010social, michael2001living}. Research has shown that social connectedness not only boosts life satisfaction among older adults but also extends their ability to live independently and supports active aging \cite{o2017definition, michael2001living, annele2019definitions}. 
Community-based social programs, such as craft groups, board games, and fitness classes, are effective approaches for older adults to establish social connections \cite{agedcareguide_social_support,lindsay2018mixed,xing2023keeping}. While beneficial, many older adults face challenges such as changes in physical health, geographic distance, or financial limitations, which make it difficult for them to engage in community events \cite{coyle2012social, pugh2009social}.
This can lead to an increased sense of loneliness and social isolation, which not only affect older people's mental health but can result in cognitive decline and deterioration of physical health \cite{house1988social,eng2002social,holt2021loneliness,nicholson2012review}. Therefore, it has become an open challenge to design assistive technologies to help older adults engage in these community-based activities and strengthen their social ties remotely through a digital connection.

Virtual reality (VR) has demonstrated the potential to help people connect with each other across geographic boundaries \cite{li2021social,zamanifard2019togetherness,freeman2022working}. VR can create immersive 3D environments that simulate face-to-face social experiences, thereby enhancing users' social presence \cite{kreijns2022social}. A range of work has leveraged VR to support healthy aging, including the use of VR as a tool to aid physical and cognitive health \cite{du2024lightsword,appel2020older,eisapour2018game,mirelman2016addition}, alleviating loneliness and mental health issues \cite{dickens2011interventions,thach2020older,baker2020evaluating}. As VR becomes more accessible, it offers older adults opportunities to enrich their social interactions in novel ways. 
However, existing social VR designs for older adults primarily focus on verbal communication, such as reminiscence conversations \cite{baker2021school} and casual chats \cite{baker2019interrogating, wei2023bridging}. While this is important for emotional connection, social bonding is not limited to verbal communication; engaging in shared activities is also an important aspect of strengthening social ties by enabling users to naturally exchange ideas and emotions while having fun \cite{evjemo2004supporting,richter2018relations,viguer2010grandparent}. 

Research suggests that culturally relevant activities are an important form of spiritual nourishment and have a positive impact on older adults' physical and mental health, social interaction and quality of life \cite{cohen2006impact,bernardo2020role}.
Inspired by this line of work, we designed a VR community to construct a social community and a culturally relevant activity (i.e., Chinese chess) for older adults to play while engaging in social interaction. We chose Chinese chess as a shared activity for older adults to participate in VR because in the Chinese context, Chinese chess (also known as Xiangqi) holds a special place in the hearts of many Chinese older adults. As a time-honored and community-based strategy game, it is not only a recreational activity, but also an important way to promote cognitive engagement and social interaction \cite{hu2012effects, shi2023effect, lin2023social, lee2018effect}. Here we proposed our first research question (RQ1): What design considerations should be taken to build a VR community that effectively supports older adults' social interaction while playing Chinese chess?

To answer RQ1, we first conducted a formative study with a total of 40 videos of older adults playing Chinese chess collected from Chinese online video platforms (e.g., YouTube, Bilibili, Douyin, Kuaishou). We analyzed these videos to extract common user behaviors and design elements from real-world scenarios, so as to explore the design of a VR chess community for elder connectedness. Subsequently, building on prior work \citep[e.g.,][]{siriaraya2019social,mcveigh2019shaping,harris2019asymmetry} and findings of our \sout{elicitation} \rv{formative} study, we derived three design considerations, including: 1) recreate familiar and immersive environmental elements for older adults, 2) provide an asymmetric Chinese chess experience, and 3) incorporate cultural elements as social catalysts. Following these design considerations, we designed RemoteChess, a multi-user VR system that provides older adults with a familiar Chinese chess experience in a virtual park. RemoteChess presents older adults with two scenes \rv{that have different social tendencies}: \textbf{\textcolor{green}{Cultural Corridor}} and \textbf{\textcolor{yellow}{Chess Table}}. In the \textbf{\textcolor{yellow}{Chess Table}} scene, asymmetric game characters \sout{experiences} (i.e., chess player and chess spectator) are introduced to enrich the social dynamics.

We conducted a user study with 18 older adults to understand whether and how RemoteChess might enhance older adults' social connectedness and gain further design opportunities to enhance social connectedness. We focus on two aspects: the enhancement of social connectedness and older adults' social behaviors in RemoteChess. Both quantitative measures and qualitative feedback were employed to answer \rv{the following two RQs:} \sout{our research questions (RQs): RQ1: Does RemoteChess enhance social connectedness and interaction among older adults? If so, RQ2: How do older adults socialize and interact with each other in RemoteChess?}
\rv{\begin{itemize}
    \item RQ2: Does RemoteChess enhance social connectedness and interaction among older adults?
    \item If so, RQ3: How do older adults socialize and interact with each other in RemoteChess?
\end{itemize}}

Our findings indicate that RemoteChess enhanced participants’ social connectedness by offering familiar environments, multi-sensory cues, incorporation of cultural elements, and asymmetric interactions. To our knowledge, we were the first to use a culturally relevant activity to build a VR social community for elder connectedness. We further discussed design guidelines for designing culturally relevant social activities in VR to promote social connectedness for older adults. To summarize, we made the following contributions:
\begin{itemize}
    \item We designed a VR social community that allows older adults to engage in a culturally relevant activity (i.e., Chinese chess) and to interact with each other. Such design offers a new perspective on enhancing older adults' social connectedness.
    \item We conducted a user study involving six groups of older adults to understand how RemoteChess promoted social interactions among older adults through playing and engaging in a culturally relevant activity.
    \item We provide design guidelines for future social VR designs aimed at social connectedness of older adults, including recreating familiar social scenes in virtual environments, incorporating cultural elements as social catalysts, and designing flexible interaction mechanisms to meet the social needs of different users.
\end{itemize}

\section{Related Work}

To clarify our research motivation and research gap, in this section, we reviewed relevant studies on social connectedness among older adults, VR to promote social connection, and background of Chinese chess and its benefits for seniors.

\subsection{Social Connectedness in Later Life}

Social connectedness, i.e. the experience of belonging and relatedness between people \cite{van2009social}, is an essential human need \cite{easton2013investigation}. As a key \rv{factor in} healthy aging, \rv{social connectedness} \rv{brings} numerous benefits to the health and well-being of older adults \cite{ashida2008differential}. Research has shown that maintaining social connections can improve life satisfaction of older adults \cite{o2017definition}, prolong their independence and integration into social life \cite{michael2001living}, and also contribute to active aging \cite{annele2019definitions}. Additionally, a strong social support network helps to strengthen the immune system and increases the chances of living a longer life \cite{holt2010social}.

Community, as an important social structure, plays a vital role in maintaining social connections and providing emotional support for older adults \cite{brossoie2003community,markle2018community,salman2021community}. Participation in community-based social programs such as craft groups, board games, square dance, and fitness classes, serves as an important way for many older people to sustain their social ties \cite{agedcareguide_social_support}. These programs offer older adults ample opportunities for social engagement, including building new friendships, obtaining social support, and fostering a sense of belonging through participation in group activities \cite{lindsay2018mixed,xing2023keeping}. By engaging in these activities, older adults can establish friendships and social support networks within their communities, which contributes to both their physical and mental well-being \cite{agedcareguide_social_support,lindsay2018mixed}.

However, not every older adult has the opportunity to participate in community-based social activities in person. Due to life course changes such as deteriorating health, retirement, and loss of intimate relationships \cite{coyle2012social,pugh2009social}, many older adults face a decline in social connectedness. This decline could further lead to increased social isolation and loneliness in later life, which are significant risk factors for health and well-being \cite{house1988social,nicholson2012review}. Studies have also shown that older adults who lack social connectedness are more likely to suffer from chronic diseases, depression, cognitive decline, and even risk premature death \cite{eng2002social,holt2021loneliness}. Therefore, there is an urgent need to investigate how to help older adults engage in these community-based activities and strengthen their social ties remotely through a digital connection. 

\subsection{VR to Support Social Connection}

Recently, the rapid development of VR has provided a new interactive platform for people to engage in meaningful social activities without geographical constraints. Social VR, as an emerging immersive remote communication tool \cite{li2021social}, allows multiple users to join a collaborative virtual environment and communicate with each other through avatars, thereby simulating face-to-face social experiences. As social VR platforms (e.g., VRChat, Rec Room, and AltspaceVR) become increasingly prevalent in the market, many people turn to these platforms to experience social connectedness. 
To date, researchers have studied various aspects of social VR experiences, including \sout{user preferences for avatar use \cite{gonzalez2018avatar},} how social VR supports mental health \cite{deighan2023social}, facilitates everyday collaborative activities \cite{freeman2022working}, understands and measures social behaviors and interactions \cite{mcveigh2018s,li2019measuring}, as well as communicating effectively in embodied environments and addressing platform governance challenges \cite{smith2018communication,blackwell2019harassment}. Researchers have also explored how social VR affects meaningful relationships \cite{zamanifard2019togetherness}, and that social VR users begin to value social VR communities and make friends easily \cite{piitulainen2022vibing}.


\sout{VR has been widely demonstrated to have benefits in promoting older adults' cognition, health, and physical functioning \cite{appel2020older,mirelman2016addition,roberts2019older,eisapour2018game,dickens2011interventions,du2024lightsword}.}
\rv{VR offers numerous potential benefits for older adults, including promoting health and physical functions \cite{appel2020older, mirelman2016addition}, positively impacting cognitive abilities \cite{appel2020older, eisapour2018game,du2024lightsword}, and improving their social and emotional well-being \cite{dickens2011interventions,roberts2019older}. Previous studies have explored VR’s usability among older adults, revealing both challenges and opportunities \cite{brown2019exploration,baker2020evaluating,seifert2021use}. Key issues include usability concerns, particularly for those with functional limitations or dementia \cite{baker2020evaluating}. Despite these challenges, VR shows promise in engaging older adults, potentially improving quality of life and social interactions \cite{brown2019exploration,baker2020evaluating,seifert2021use}.}
Recently, researchers began to explore how VR can be used by older adults to sustain their social connections.
One of the advantages of VR interventions is that they allow older adults, who may have physical limitations or live far away, to safely and comfortably participate in meaningful social activities from the comfort of their own homes. Baker et al. reported a series of studies investigating how older adults communicate with each other in VR \cite{baker2019interrogating}, how avatars in social VR meet the communication needs of older adults \cite{baker2021avatar}, and how social VR can support group reminiscence among seniors \cite{baker2021school,baker2019exploring}. Their work shows that VR can not only provide immersive social experiences, but also help older adults build and maintain social connections, effectively reducing feelings of loneliness and social isolation. Similarly, research by Afifi et al. indicated that VR can serve as a bridge, bringing older adults closer to family members \cite{afifi2023using} and fostering inter-generational communication and emotional bonding \cite{wei2023bridging}. These studies highlight the importance of using VR to foster reciprocal and enjoyable social connections among older adults, suggesting that social VR has great potential for application in future aging society.

Although VR shows great potential in enhancing social connectedness, most existing studies focus on general user groups. There is still a lack of research on specifically designing social VR experiences to enhance social connections among older adults. Furthermore, the few existing social VR platforms designed for older adults tend to prioritize verbal communication, such as reminiscence conversations \cite{baker2021school} and casual chatting \cite{baker2019interrogating}. While verbal interaction plays a significant role in fostering emotional connection, social bonding extends beyond just conversation. Participation in shared activities is another key factor in strengthening relationships. Research indicates that participating in shared activities allows individuals to naturally share ideas and emotions while having fun together \cite{evjemo2004supporting,richter2018relations,viguer2010grandparent}. A study by Wei et al. has highlighted the need for social VR platforms to move beyond solely communication-based interactions and instead offer a variety of shared experiences to promote deeper social connections \cite{wei2023bridging}.
\sout{In this work} \rv{As VR becomes more accessible}, we sought to find a shared activity familiar to older adults and construct a \rv{VR social} community \sout{in VR}. By leveraging the social benefits of both VR and shared activities, we hope to provide a unique platform for older adults to maintain and strengthen their social ties, thereby contributing to their overall well-being.

\subsection{Chinese Chess \sout{(Chinese chess)} Among Older Adults}

Chinese chess is a traditional strategic board game that represents a battle between two players \cite{leventhal1978}. The game is played on a flat board with pieces divided into two sets, typically red and black, each set consisting of 7 different Chinese characters and 16 pieces in total \cite{lau2011chinese}. Participants need to capture the enemy's general (king). In China, Chinese chess is a highly popular form of social entertainment among older adults, as it offers a familiar and intellectually stimulating pastime. In order to win or play well in the game, players need to coordinate and work with various abilities such as attention, memory, logical thinking, and decision-making. Repeated use of these abilities during gameplay can help maintain and improve cognitive functioning in older adults \cite{hu2012effects,shi2023effect}. Nuria et al. \cite{cibeira2021effectiveness} suggest that there is a significant correlation between regular board gameplay and a better quality of older life. Moreover, the social aspect of the game is equally significant; Chinese chess facilitates social interaction, providing a platform for emotional support and community bonding among seniors \cite{lin2023social,lee2018effect}.

Despite the benefits, not every older person has the opportunity to participate in such activities in an outdoor environment due to poor health and disabilities, geographic distance, or limited income. This prevents them from maintaining social networks and developing cognitive exercises. Various digital platforms and technologies have been developed to enable remote chess gameplay in previous research and market, such as computer or mobile chess games \cite{qqchess_webpage,bontchev2008mobile}, chess robots \cite{larregay2018design,kolosowski2020collaborative}, augmented reality (AR) applications \cite{chen2008remote,cerron2023multiplayer,yusof2019collaborative}, and VR platforms \cite{very_real_chess_steam,chess_vr_multiverse_journey_steam,vrchat_chess_world}. However, these existing systems tend to offer generalized social experiences that may not resonate with older adults. They often fail to replicate the rich social dynamics of face-to-face Chinese chess gatherings, which include not only the game itself but also social chit-chat, story sharing, and cultural exchange.

\section{Design of RemoteChess}


\sout{To inform the design of a VR community for older adults using Chinese chess as a medium, we conducted an elicitation study (Sec. \ref{sec:formativestudy}) to understand how older adults engage in social interactions while playing Chinese chess.}
\rv{To answer RQ1, we conducted a formative study (Section \ref{sec:formativestudy}) to understand how older adults engage in social interactions while playing Chinese chess, so as to inform the design of a VR community for older adults using Chinese chess as a medium.} Then, we identified three design considerations (\rv{Section} \ref{sec:dcs}) based on the findings from our \sout{elicitation} \rv{formative} study and previous literature in the HCI field. These considerations further guided our design of an immersive VR experience that is both socially and culturally engaging for older adults.   

\subsection{\sout{Elicitation} \rv{Formative} Study}
\label{sec:formativestudy}

While Chinese chess originated in China and has a large number of enthusiasts there, it has also gained many fans in other countries with the growing trade and cultural exchanges. Therefore, to cover a wide range of geographic regions and cultural backgrounds, we chose to study the social interaction patterns of older adults playing Chinese chess by analyzing videos on social media. \rv{This method of} video analysis \cite{knoblauch2012video, vom2016action} is non-intrusive, \rv{as it} does not directly affect the subjects being observed, allowing us to capture \rv{older adults'} genuine behaviors in a more natural way.

\subsubsection{Data Collection}

In order to collect videos of older adults playing Chinese chess in real-life scenarios, we searched on YouTube, the world's largest video search and sharing platform, using keywords related to older adults playing Chinese chess (e.g., older adults playing Chinese chess, senior citizens playing Chinese chess, and retirees playing Chinese chess). Since there is a relatively large Chinese chess audience in China, we also used the same keywords to search on major video platforms in China (e.g., Bilibili, Douyin, Kuaishou, etc.). We filtered the results based on relevance, paying special attention to videos that included interaction, conversations, and detailed behaviors. In total, we collected 40 videos and organized their links into \rv{an Excel} table.

\subsubsection{Observation and Data Analysis}

We formulated a coding framework \rv{in NVivo\cite{nvivo2024}} to document our observations, covering the following aspects: Behavior, Communication, Environment, and Others (any other possible observations). Then, three researchers independently watched and analyzed the videos listed in the table, recording their observations according to the consistent coding framework \rv{in NVivo}. Researchers took detailed notes on the content of each video, people's social behaviors, environmental elements, and so on. \rv{Afterwards, the three researchers shared their analysis results.} \sout{Afterwards,} We summarized and compared the observations of the three researchers, identifying commonalities and differences to ensure the completeness and objectivity of the data.

\subsubsection{Findings} By analyzing videos of older adults playing Chinese chess\sout{in real-life scenarios}, our key findings are as follows:

\begin{figure*}
    \centering
    \includegraphics[width=\textwidth]{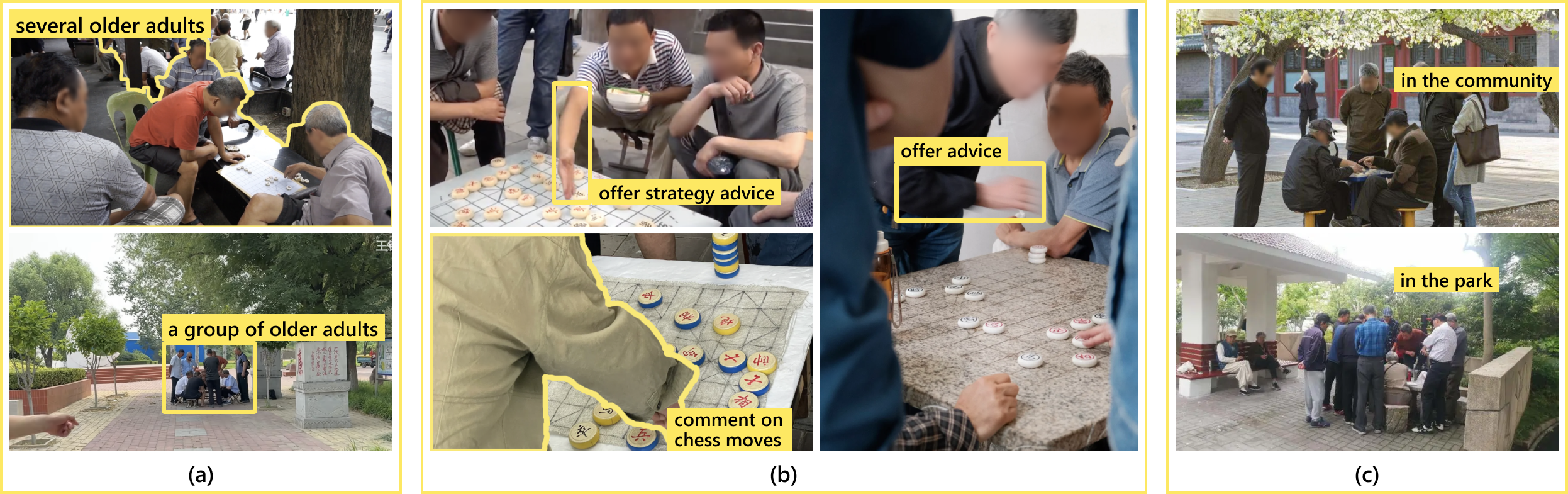}
    \caption{Example scenes from online videos showing the patterns and preferences of older adults while playing Chinese chess: (a) Older adults gather in outdoor environments (e.g., public parks) to have Chinese chess experience; (b) Chess spectators are always trying to help; (c) Places where older adults play Chinese chess are usually in open outdoor environments. }
    \Description{Example scenes from online videos showing the patterns and preferences of older adults while playing Chinese chess: (a) Older adults gather in outdoor environments (e.g., public parks) to have Chinese chess experience; (b) Chess spectators are always trying to help; (c) Places where older adults play Chinese chess are usually in open outdoor environments.}
    \label{fig:formative}
\end{figure*}

\textbf{Older adults often play Chinese chess in groups.} Our video analysis results show that 36 out of 40 videos show older adults playing Chinese chess in groups. We found that Chinese chess playing among older adults has a social nature of group activities (Figure \ref{fig:formative}(a)). Although there are only two players in each game, the spectators usually gather around, creating a lively social atmosphere. In such a setting, spectators do not just passively watch the game, but also actively participate in discussions and express their opinions. \sout{Videos show that older adults not only focus on the game itself, but also engage in a great deal of chit-chat and communication during the game. This type of social interaction is an important part of their Chinese chess experience.} \rv{Additionally, videos show that older adults not only focus on the strategy and advice conversations around the game itself, but also engage in some life-related small talks during the game. This type of social interaction with a wide range of conversations is an important part of older adults' Chinese chess experiences.}

\textbf{Active participation of chess spectators.} We observed a variety of roles in the videos, including game players, quiet spectators, and active advisors. This asymmetric distribution of roles increases the complexity of social interactions and provides more opportunities and ways for people to participate. Although there is a Chinese proverb that says, "a true gentleman remains silent while watching a chess game", it is shown in the videos that many chess spectators are always trying to help the players by offering strategy advice or commenting on chess moves (Figure \ref{fig:formative}(b)). This type of behavior suggests that most spectators are not just spectators; they actively engage with the game in another interactive way. Such active involvement transforms Chinese chess playing from a competition between two \sout{people} \rv{individuals to} a more social activity, \sout{which promotes mor} \rv{thus facilitating the inclusion and interaction of more participants.}

\textbf{Preferences for open and natural environment settings.} Through the videos, we observed that older adults often play Chinese chess in outdoor settings (Figure \ref{fig:formative}(c)), such as the shade of parks, inside gazebos, and on neighborhood streets. These venues are common gathering spots for older adults, providing a familiar and comfortable atmosphere that helps them to relax and communicate freely. This choice of environment reflects older adults’ preferences for open and natural spaces.

\subsection{Design Considerations}
\label{sec:dcs}

Drawing from the findings of our \sout{elicitation} \rv{formative} study and previous literature on social VR design, we identified the following design considerations to guide the design and development of RemoteChess. 

\textbf{DC1: Resemble Familiar Environmental Elements for Older Adults in VR.}
When designing a VR social community for older adults, our primary goal is to promote greater social presence \cite{biocca2003toward,kreijns2022social} and immersion within the environment. Prior work emphasized the importance of using familiar real-world settings to shape behavioral expectations in VR \cite{mcveigh2019shaping}. McVeigh-Schultz et al. mentioned that locations in social VR are often designed to leverage social expectations from the physical world; they also suggested evoking a sense of familiar social environments through the aesthetics of places and architecture \cite{mcveigh2019shaping}. This approach aligns with Siriaraya et al.'s findings - virtual environment design affects the quality of social interaction experienced by older adults in 3D virtual worlds \cite{siriaraya2019social}. Additionally, Yang et al. \cite{yang2020effects} explored the impact of auditory cues and found that spatial audio significantly enhances the feeling of being in the same space as others. 
Informed by these studies, RemoteChess uses an outdoor park as its virtual environment, as the park is where older adults often gather to play Chinese chess (Figure \ref{fig:formative}(c)). To resemble the environment’s characteristics and enhance the immersion of older adults, RemoteChess adds multi-sensory elements such as environmental sounds that are familiar to them when playing Chinese chess.

\textbf{DC2: Provide asymmetric Chinese chess experience.}
Prior work revealed that social presence in VR is influenced by asymmetric interactions \cite{yassien2020design}. In our \sout{elicitation} \rv{formative} study, we found that although the game of Chinese chess itself is symmetric, the surrounding social interactions are often asymmetric, such as different roles and participation of players and spectators in the game (Figure~\ref{fig:formative}(b)). This observation aligns with Harris et al.’s findings, who noted that compared to symmetric games, asymmetric game experiences can enhance participants' social connection, behavioral engagement, and immersion \cite{harris2019asymmetry}. Therefore, RemoteChess incorporates different roles such as “chess player” and “chess spectator”, so as to simulate this kind of asymmetric interaction and encourage a wider range of social interactions.

\textbf{DC3: Incorporate cultural elements as social catalysts.}
Chinese chess is not only a recreational activity, but also contains rich cultural significance which can serve as a bridge for discussion among people. Prior work supports positioning social catalysts and other attention focusing objects in VR to serve as a social lubricant, sparking conversation and interaction among users \cite{mcveigh2019shaping}.  McVeigh-Schultz et al. suggested taking advantage of social contagion, such as cultural practices associated with community outreach, to promote social interaction in social VR \cite{mcveigh2019shaping}. Inspired by this, RemoteChess incorporates Chinese chess history display and related cultural elements in the VR environment. These elements act as cues and topics for people’s social interaction, providing older users with opportunities for discussion and sharing, thereby enhancing their social engagement and sense of belonging to the community.

\rv{\subsection{Iterative Design Process}

During the iterative design process of the VR system, researchers began by brainstorming and sketching concepts to explore various possibilities for the virtual environment and user interactions. Then one researcher started developing more detailed wireframes and prototypes, trying out various layouts for the virtual park and user interfaces. We held discussions every two weeks to review each version of the prototype, revised the design based on feedback from other team members. 
Additionally, we chose a simplified avatar style for older adults' virtual representations. Because previous study suggests that avatars are not considered directly important for older adults' social interaction in 3D virtual environments, but rather seen as a "placeholder" to complete tasks \cite{siriaraya2019social}.
}

\section{The RemoteChess Community}

Based on the explored design considerations, we developed the RemoteChess system, which presents older adults with two scenes: \textbf{\textcolor{green}{Cultural Corridor}} and \textbf{\textcolor{yellow}{Chess Table}}.

\begin{figure*}
    \centering
    \includegraphics[width=0.99\textwidth]{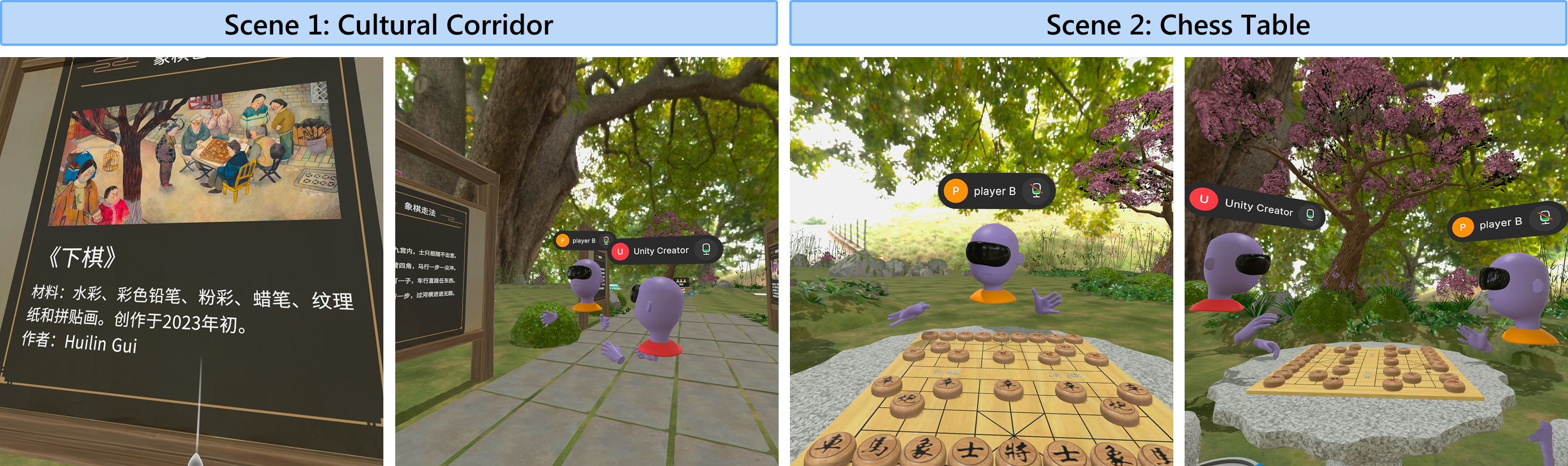}
    \caption{This figure shows two main scenes of RemoteChess. (a) An exhibition of  Chinese chess-related painting in Scene 1; (b) Two users watching and discussing the cultural exhibition; (c) Chess player’s view; (d) Chess spectator’s view.}
    \Description{This figure shows two main scenes of RemoteChess. (a) An exhibition of  Chinese chess-related painting in Scene 1; (b) Two users watching and discussing the cultural exhibition; (c) Chess player’s view; (d) Chess spectator’s view.}
    \label{fig:enter-label}
\end{figure*}

\begin{figure*}
    \centering
    \includegraphics[width=\textwidth]{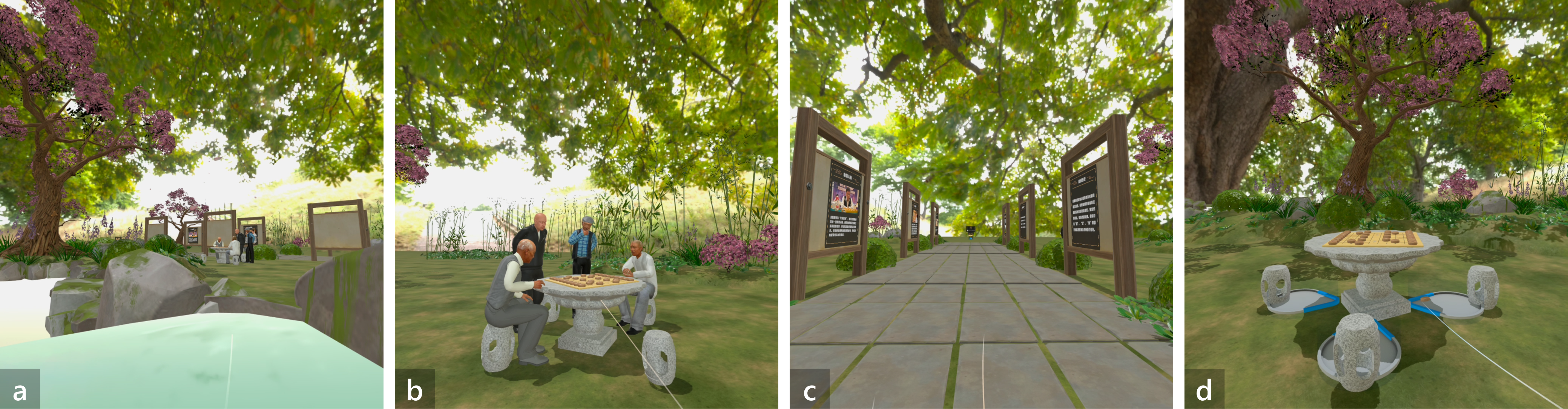}
    \caption{Some example scenes of RemoteChess VR community. (a) The virtual environment of a park; (b) Non-interactive Chinese chess avatars in the park; (c) Scene 1 - cultural corridor; (d) Scene 2 - chess table.}
    \Description{Some example scenes of RemoteChess VR community. (a) The virtual environment of a park; (b) Non-interactive Chinese chess avatars in the park; (c) Scene 1 - cultural corridor; (d) Scene 2 - chess table.}
    \label{fig:enter}
\end{figure*}

\textbf{Virtual Environment } The RemoteChess community emulates an outdoor park setting, a common location for older adults to gather and play Chinese Chess in real life (\textbf{DC1}). This virtual environment incorporates familiar elements such as natural scenery and ambient sounds like birdsong and rustling leaves, creating a multi-sensory experience that enhances comfort, immersion, and social interaction.

\textbf{Scene 1 - \textcolor{green}{Cultural Corridor} }
In Scene 1, we utilized a series of display contents related to Chinese chess history, culture, and art for older adults to explore. This scene not only provides background information on Chinese chess culture, but also provides topics and discussion points for interaction between users through exhibition design (\textbf{DC3}). By adding cultural elements related to Chinese chess, we aim to make Scene 1 a catalyst for interaction between users, helping them to establish deeper social connections in a common cultural context.

\textbf{Scene 2 - \textcolor{yellow}{Chess Table} }
In Scene 2, we positioned two different roles for users to choose from: chess players and chess spectators (\textbf{DC2}). This asymmetric role allocation simulates the social dynamics of older adults playing Chinese chess in real life. Not only can chess players actively participate, but spectators can also participate in the game process through discussions and suggestions. We hope that through this design, we can enhance the diversity and fun of social interaction, so that more users can find a way of participation that suits them and meet different social needs.

\rv{\textbf{Two scenes with different social tendencies} To better meet the diverse socialization needs of older adults, we constructed two scenes with different social tendencies. Scene 1 is a topic-oriented social scenario, whose main goal is to promote conversations and interactions through common interests (culture and history of Chinese chess). We provide users with a common topic starting point: they can start a discussion around the displayed cultural content. Scene 2, on the other hand, is an activity-oriented social scenario that focuses on simulating real-life social dynamics through Chinese chess games.}

\textbf{Implementation } We developed RemoteChess community using Unity 3D (2022.3.20f1c1), XR Interaction Toolkit, leveraging Unity’s VR Multiplayer template \cite{unity_vr_multiplayer_template} for multiplayer network interaction and voice chat.

\section{User Study}

\begin{table*}[h]
    \centering
    \caption{Participants' demographic information.
    We used the following terms to represent their Chinese chess level: \textbf{low} (basic understanding with limited skill), \textbf{medium} (some experience, familiar with rules), and \textbf{high} (extensive experience, skilled in complex strategies). Each participant's level was determined based on their self-reported experience during the background interview.}
    \label{tab:participants}
    \footnotesize
    \begin{tabular}{lcccccc} 
    \toprule
    & ID & Age & Gender & Prior VR Experience & Chinese Chess Level & Social Relations \\ 
    \midrule
    & P1 & 77 & M & no & high & \\
    Group 1 & P2 & 76 & F & no & medium & P1, 2 are couples \\
    & P3 & 60 & F & no & low & \\
    \midrule
    & P4 & 60 & M & yes & low & \\
    Group 2 & P5 & 65 & F & no & low & P5, 6 are couples \\
    & P6 & 80 & M & no & low & \\
    \midrule
    & P7 & 63 & F & no & medium & \\
    Group 3 & P8 & 66 & M & yes & high & P7, 8 are couples \\
    & P9 & 63 & F & no & low & \\
    \midrule
    & P10 & 62 & F & no & medium & \\
    Group 4 & P11 & 68 & F & no & medium & strangers \\
    & P12 & 61 & F & no & low & \\
    \midrule
    & P13 & 75 & M & no & low & \\
    Group 5 & P14 & 61 & M & no & low & P14, 15 are friends \\
    & P15 & 60 & F & no & low & \\
    \midrule
    & P16 & 76 & M & no & medium & \\
    Group 6 & P17 & 62 & F & no & low & strangers \\
    & P18 & 68 & M & no & low & \\
    \bottomrule
    \end{tabular}
\end{table*}

To understand whether and how RemoteChess might enhance social connectedness among older adults, we conducted a user study to assess their experience and the usability of this platform. We focused on two aspects: 1) the enhancement of social connectedness and interaction; and 2) older adults' social behavior patterns in RemoteChess. The first aspect responds to \textbf{RQ2}, and the second responds to \textbf{RQ3}. \sout{Before starting the study, our university ethics committee (IRB) had granted ethical approval.} \rv{The study was approved by our university's ethics committee (IRB).}

\subsection{Participants}

We recruited 18 older adults (i.e., 10 females, 8 males; age: M=66.83 years, SD=6.68 years) from a local senior college to participate in the study.
Participants’ demographic information is detailed in Table~\ref{tab:participants}.
All participants had experience playing Chinese chess and showed interest in this activity.
To ensure the smooth running of our experiment, participants were required to meet the following health criteria: 1) no serious vision problems; 2) no mobility disorders; and 3) no tendency towards motion sickness.
All participants were paid after completing the experiment.

\subsection{Apparatus}

As for the devices, we used two Meta Quest Pro \cite{metaQuestPro} and one Meta Quest 2 \cite{metaQuest2} for the experiment. Participants interacted with the VR system using two hand controllers to teleport within the scene and click on user interfaces. When playing Chinese chess, participants used bare hands to interact with chess pieces to replicate the real-life experience.

\subsection{Procedure}

\begin{figure*}
    \centering
    \includegraphics[width=\textwidth]{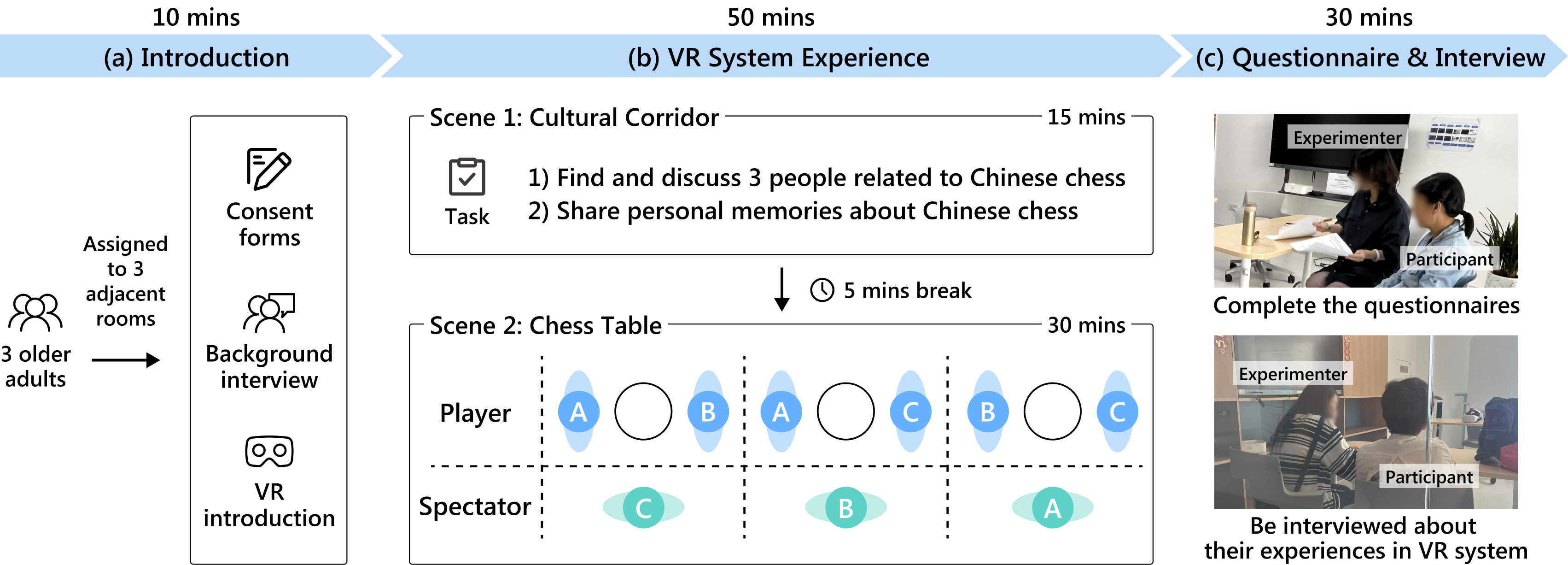}
    \caption{User study procedure. (a) Be assigned to three rooms for background interviews and VR introduction. (b) Experience two VR scenes in RemoteChess and interact with other participants. (c) Complete two questionnaires and be interviewed about experiences and perceptions in RemoteChess.}
    \Description{User study procedure. (a) Be assigned to three rooms for background interviews and VR introduction. (b) Experience two VR scenes in RemoteChess and interact with other participants. (c) Complete two questionnaires and be interviewed about experiences and perceptions in RemoteChess.}
    \label{fig:procedure}
\end{figure*}

We recruited a total of 6 groups of older adults, with 3 people in each group. Since RemoteChess is a VR social community that supports multiple older users participating remotely, we assigned each group's participants to 3 separate rooms to experience VR communication and simulate remote conditions. Each room was equipped with an experimenter, a VR headset, and a laptop.
Each VR headset's viewport was projected onto a laptop using Quest Link in case of technical issues.
We used Tencent Meeting (a video conferencing application) to conduct and record all study sessions across the three rooms. Additionally, we set up three cameras in the experimental area to capture the participants' physical movements in each room. The study lasted about 90 minutes and consisted of three parts (Figure~\ref{fig:procedure}).

\textbf{Part 1: Background Interview and VR Introduction.} To begin with, we introduced our user study process to the participants and had them fill out consent forms. We conducted a background interview to gather information about their age, social networks, experience with Chinese Chess, prior usage of VR devices, and whether they have ever felt lonely or socially isolated in their daily lives. Next, the experimenters introduced VR technology and helped participants learn how to use the VR headsets for basic operations. Before the formal experiment began, participants practiced simple tasks using the controllers, such as moving in the virtual environment and grabbing and placing objects. This session lasted 10 minutes and aimed to ensure all participants were comfortable using the VR equipment and to enhance their operational skills. Participants could ask any questions about VR during this part.

\textbf{Part 2: VR System Experience.} In this session, participants experienced two scenes of RemoteChess: \textbf{\textcolor{green}{Cultural Corridor}} (Scene 1) and \textbf{\textcolor{yellow}{Chess Table}} (Scene 2). Considering their limited experience with VR, experimenters provided guidance and support to facilitate participant interaction. First, participants entered Scene 1 to explore the history, culture, and art related to Chinese chess. To encourage discussion, experimenters provided two tasks for the participants: 1) discuss three people displayed in the corridor and their influence on Chinese chess, and 2) share personal stories or memories related to Chinese chess. Each task took around 5 minutes. Then participants had a 5-minute break, during which they removed the VR headsets to alleviate any possible visual fatigue from using VR.
Afterward, participants put the VR headsets back on and entered Scene 2.
To ensure everyone had the chance to experience both the role of chess player and chess spectator, participants had three rounds of Chinese chess, each for 10 minutes. During the whole session, we encouraged participants to share their understanding of the game and engage in natural social interactions. 

\textbf{Part 3: Questionnaire and Semi-structured Interview.} After experiencing the VR system, participants completed two questionnaires. 
Then three participants were interviewed separately for 20-30 minutes.
This part sought to gain a deeper understanding of participants' experiences and perceptions of RemoteChess.

\subsection{Measurements}

We collected both quantitative and qualitative data to assess participants’ experiences within RemoteChess. Quantitative data includes two questionnaires and participants' interaction data. All items of the questionnaires were administered on a 5-point Likert scale (1: strongly disagree, 5: strongly agree). To assess the overall usability of RemoteChess, we chose the System Usability Scale (SUS) \cite{lewis2018system}, which consists of ten items. We then used the Social Connectedness Scale (SCS) \cite{carroll2017conceptualization} to measure participants’ perceived social connectedness. Participants rated their feelings of social connectedness across Scene 1 and Scene 2 using three dimensions: inclusion of others, belonging to others, and comfort with others.
Also, we collected interaction data of our participants within RemoteChess, including verbal turns, time spoken, and words spoken. By analyzing these aspects, we can measure different levels of engagement of our participants in the two scenes and gain insights into how RemoteChess promotes older adults' social interactions.

\sout{As for qualitative data, we focused on the social behaviors exhibited by participants in RemoteChess and their interview responses.} \rv{As for qualitative data, we focused on participants' conversation content and social behaviors in RemoteChess, as well as their interview responses.} Social behaviors broadly refer to interactions between individuals, particularly those involving verbal communication or body languages. \sout{Three researchers independently watched the video recordings of our user study and documented the social behaviors observed between participants. Then we then discussed and summarized our findings together.}
During semi-structured interview, we asked participants about their experiences\sout{in RemoteChess} and how they felt connected to others while using the system. Example interview questions included: \textit{what aspects of RemoteChess do you like or dislike and why? which features help you connect with other users? when do you feel less connected to other users? which role did you prefer, chess player or chess spectator, and why?} Participants \sout{could} also expressed their VR community experiences by comparing their experiences with face-to-face or remote gaming systems.
\rv{The video recordings of participants' VR experiences and interviews were first transcribed into a text script. Then three researchers independently coded the script using an open-coding approach \cite{corbin2014basics}. Following a weekly discussion, we shared individual coding results, achieved consensus on the final coding result, and discussed our findings together.}

\section{Results \& Findings}

In the following, we report the results of our user study. We introduced how RemoteChess enhanced older adults' social connectedness. Also, we gained a deeper understanding of how older adults use RemoteChess for socializing.

\subsection{Enhancement of Social Connectedness and Interaction \rv{(RQ2)}}

\subsubsection{Social Connectedness}
\textbf{Inclusion of Others} (whether people feel accepted and included in their social relationships). As our data followed a normal distribution using the Shapiro-Wilk test, we conducted a paired-samples t-test to assess the effect of two scenes with different social tendencies on participants' scores. The average scores for inclusion of others was M=3.67 (SD=0.49) for Scene 1, and M=4.11 (SD=0.32) for Scene 2. The analysis did not yield a significant difference between Scene 1 and Scene 2 (t(17)=-2.0456, p>0.05, Cohen's d=0.4822).
\textbf{Belonging to Others} (whether people feel that they belong to a group or social network). The perceived belonging to others was M=4.11 (SD=0.33) for Scene 1 and M=3.94 (SD=0.31) for Scene 2. However, the statistical analysis indicated that there were no significant differences between the two scenes on this sub-scale's scores (t(17)=0.7177, p>0.05, Cohen's d=0.1692).
\textbf{Comfort with Others} (how natural and relaxed people feel when interacting with others). The analysis revealed a statistically significant difference in participants' comfort level with others between the two scenes (t(17)=-4.7610, p<0.05, Cohen's d=1.1222). The comfort level participants experienced with others in Scene 2 (M=4.44, SD=0.51) was significantly higher than that in Scene 1 (M=3.55, SD=0.52). This suggests that the more familiar and interactive environment of Scene 2 may have allowed participants to feel more comfortable and comfortable interacting with others.

\subsubsection{Interaction Data}

\rv{\begin{figure*}
    \centering
    \includegraphics[width=\textwidth]{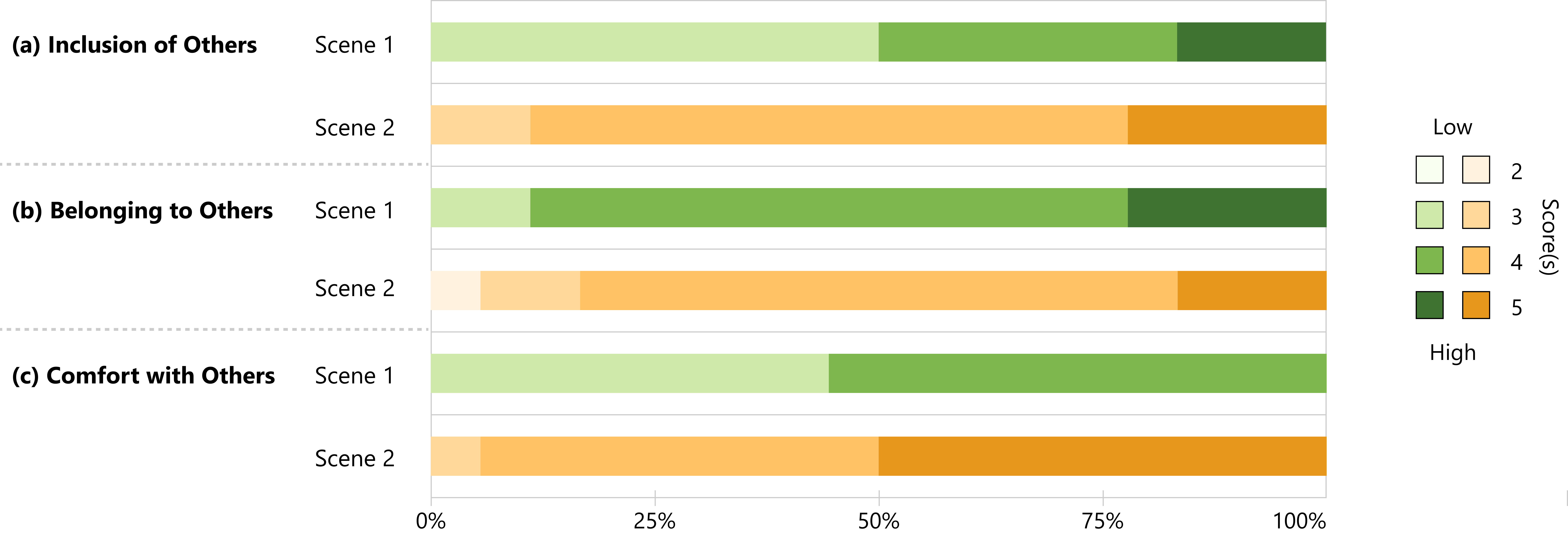}
    \caption{\sout{Participants’ ratings of the Social Connectedness Scale (SCS) based on honrizontal stacked bar charts.} \rv{Participants rated the Social Connectedness Scale (SCS) across Scene 1 and Scene 2 using three dimensions: (a) inclusion of others, (b) belonging to others, and (c) comfort with others. This figure confirmed that RemoteChess enhanced their overall social connectedness. Each subplot is composed of a horizontal bar chart to show the percentages of each rating score.}}
    \Description{Participants rated the Social Connectedness Scale (SCS) across Scene 1 and Scene 2 using three dimensions: (a) inclusion of others, (b) belonging to others, and (c) comfort with others. This figure confirmed that RemoteChess enhanced their overall social connectedness. Each subplot is composed of a horizontal bar chart to show the percentages of each rating score.}
    \label{fig:sus}
\end{figure*}}

\begin{figure*}
    \centering
    \includegraphics[width=\textwidth]{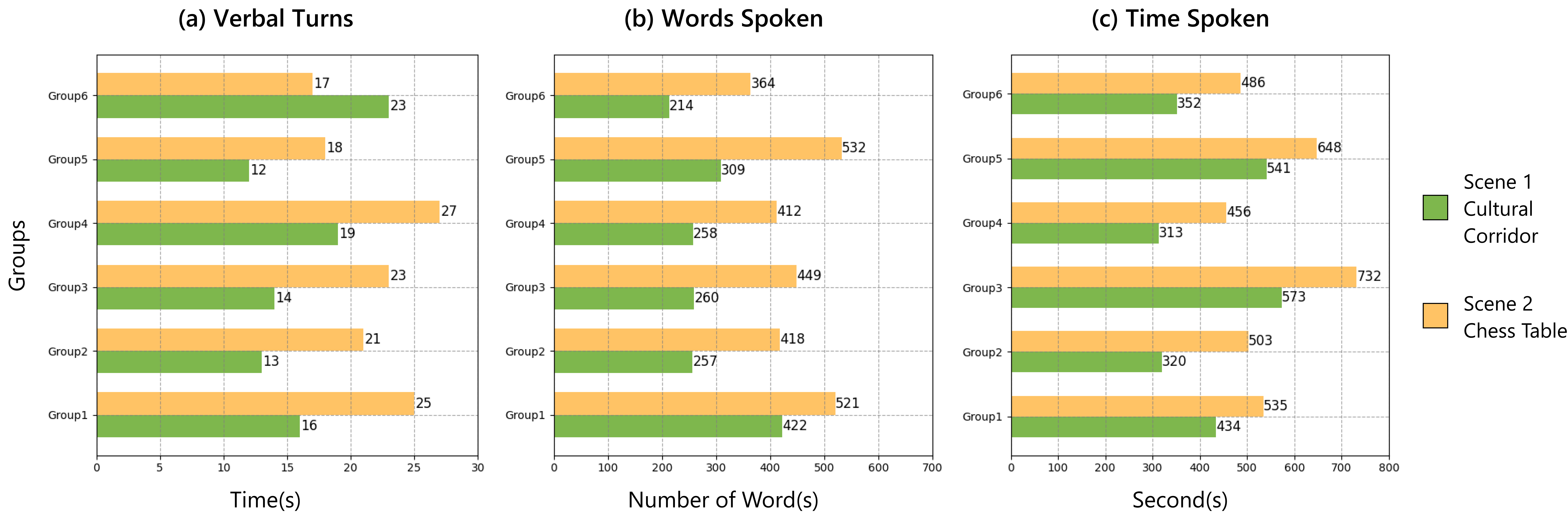}
    \caption{The six groups’ interaction data in our user study based on bar charts, including (a) verbal turns, (b) words spoken and (c) time spoken.}
    \Description{Quantitative user study results. The six groups’ interaction data in our user study based on column charts, including verbal turns, time spoken and words spoken.}
    \label{fig:data}
\end{figure*}

We counted each group’s participants’ verbal turns, time spoken, and words spoken across two scenes in RemoteChess. We then used \sout{column diagrams} \rv{bar charts} to illustrate these interaction data in Figure~\ref{fig:data}. The t-test indicated no statistically significant differences in participants' verbal turns across the two scenes (t(5)=-2.3852, p>0.05, Cohen's d=0.9737). However, results also suggested that participants' words spoken (t(5)=-9.5934, p<0.05, Cohen's d=3.9165) and time spoken (t(5)=-10.8551, p<0.05, Cohen's d=4.4316) were significantly higher in Scene 2 compared to Scene 1. This suggests that Chess Table fostered greater social interaction and engagement.

\subsubsection{Environmental elements similar to real situations facilitate social interaction}
The majority of our participants (N=15) appreciated the scene design in RemoteChess, as they found it beautiful. P2 mentioned: \textit{"The scenery is quite beautiful. Although it’s not a real park, I feel like being in a painting."} Six participants expressed that the park environment made them feel familiar and relaxed while playing Chinese chess. P9 said: \textit{"I often play Chinese chess with my friends in the park. Playing Chinese chess in this virtual park really felt like being back in a place we often visit." } This sense of familiarity helped P9 feel more immersed and willing to socialize with other participants.
Particularly, participants (N=5) highlighted the environmental sound effects, noting that these elements enhanced their sense of immersion. P10 said: \textit{"Hearing the birds and seeing the trees in the park made me feel like I was playing chess outdoors with friends." } P5 elaborated: \textit{"The natural sounds in the environment, like flowing water, cicadas, and birds, were soothing and comforting."} However, some felt that the environment was not so realistic. \textit{"The scene doesn’t feel real, like the bushes,"} P4 reported. A few participants also expressed dissatisfaction with the lack of direction in the virtual park. \textit{"Though there are two scenes, I’d like to roam around the park with my partners. The scene needs direction guidance."} P8 remarked.

\subsubsection{Cultural elements as social catalysts enhance community feeling}
Scene 1 introduced a historical exhibition of Chinese chess. This scene received generally positive feedback from our participants (N=12). Overall, participants indicated that the culturally relevant content not only triggered their discussions, but also provided opportunities for their cognitive stimulation. Additionally, participants mentioned feeling a sense of community belonging and took the opportunity to share personal stories and memories.

Participants (N=8) noted that Chinese chess culture provided rich topics for their discussion, which increased the frequency and depth of their interactions. P9 said: \textit{"The Cultural Corridor gave me a starting point for conversations, helping me find topics to talk about with others."} This highlights how traditional cultural elements like Chinese chess can serve as tools to enhance social interaction. Some participants (N=3) also mentioned that discussing Chinese chess rules, history, and techniques made it easier for them to connect with others. Several participants (N=5) shared that they learned something new by visiting Scene 1. P4 appreciated that Chinese chess culture allowed him to obtain new knowledge. P9 also found the content \textit{"really helpful"}, learning a lot about the history and culture of Chinese chess. P7 suggested: \textit{"I hope the Cultural Corridor can be updated regularly. If the exhibition remains the same, people may lose interest over time."}

Participants also mentioned that cultural elements gave them a stronger sense of community. For example, P18 felt that Scene 1 was an essential part of the community, as it fostered a sense of belonging through shared cultural themes. Some participants (N=6) felt that the Cultural Corridor evoked memories of shared history and culture, naturally leading to conversations. P10 said: \textit{"Chinese chess is not just a game. It’s a common topic for us. We find a lot of resonance when we talk."} The cultural background of Chinese chess became a bridge for deeper communication among older adults, strengthening their emotional connections. P9 shared: \textit{"Seeing the Chinese chess history exhibit made me reminisce about learning chess from my grandfather when I was young."} P17 recalled: \textit{"It reminded me of the days when I used to play Chinese chess with friends on the streets, discussing various strategies together. It felt like returning to that carefree time."}

\subsubsection{Multi-sensory experience enhances engagement}
The multi-sensory elements introduced in RemoteChess, such as environmental sounds and visual effects like leaves swaying in the wind, received positive feedback from our participants (N=13). P6, P13, and P14 all noted that these features significantly enhanced the sense of immersion, making the experience feel more realistic. P6 shared: \textit{"Hearing birds chirping and leaves rustling in the background made me feel like I was really playing Chinese chess in a park with friends."} This suggests that the natural sounds made the virtual environment more vivid and helped participants immerse themselves in the game.
Some participants expressed a desire for more sensory elements in the future. For example, P12 suggested: \textit{"It would be great if there were more sensory features, like feeling a handshake or the texture of the chess pieces. It would make the experience more real and fun."} This feedback indicates that participants hope for more tactile feedback in virtual environments, which could deepen their interaction with virtual objects.

In addition to tactile feedback, some participants emphasized the importance of sensing the presence of others. P2 said: \textit{"Physical contact is important in social interactions. I’d like to feel the warmth of human contact, like a handshake or a hug."} This desire for a "human warmth" experience reflects participants’ wish for more authentic social interactions in virtual environments, where they can feel physical touch between people. Similarly, P12 mentioned: \textit{"If you could feel a handshake in the virtual world, it would make the experience more complete."} These comments show that participants’ expectations for multi-sensory experiences go beyond enhancing sight and sound. They hope for the inclusion of tactile sensations and other sensory systems to make virtual social interactions more engaging.

\subsection{Older Adults' \sout{Social Behavior} \rv{Socializing} in RemoteChess \rv{(RQ3)}}

\begin{figure*}
    \centering
    \includegraphics[width=\textwidth]{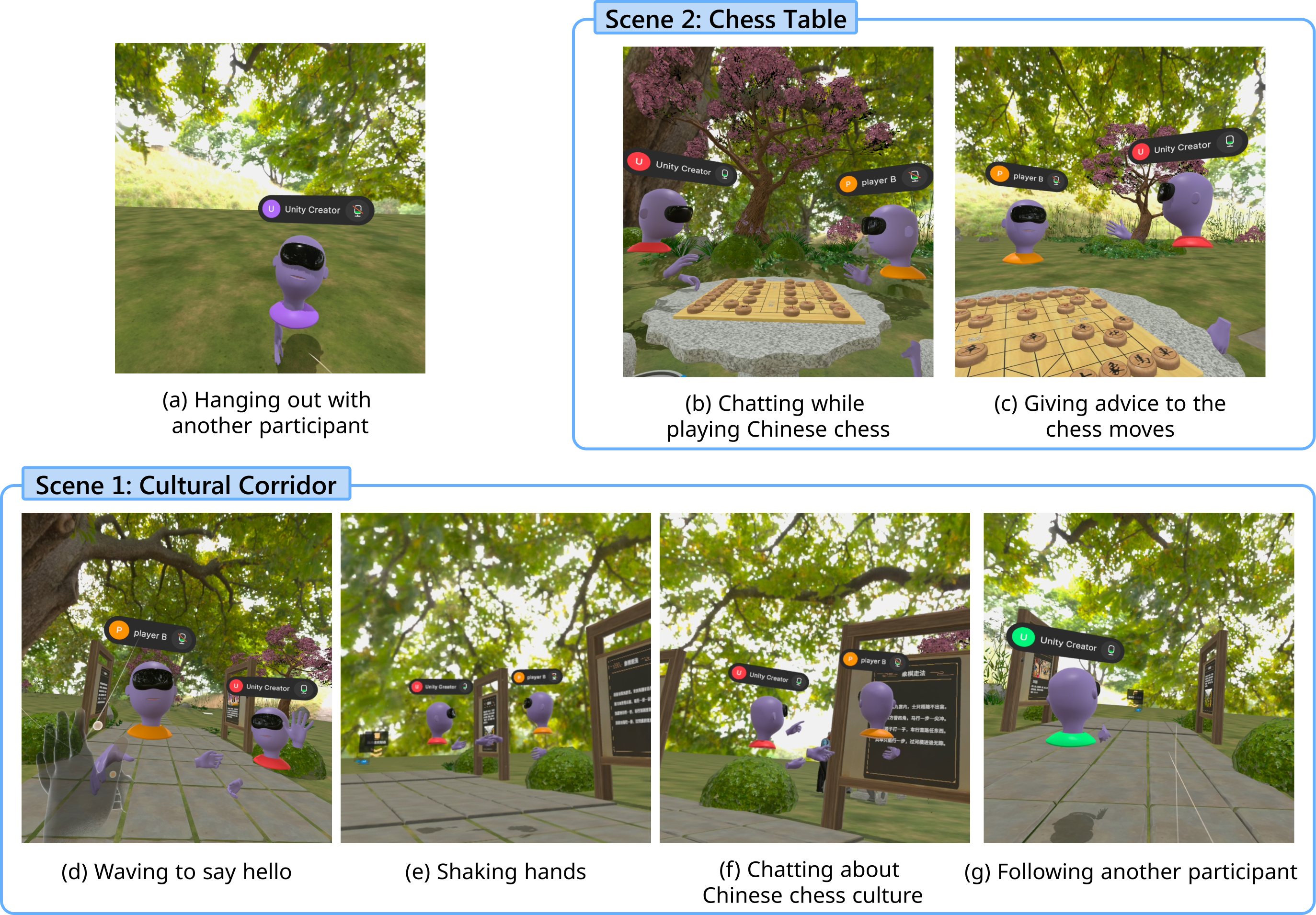}
    \caption{Screenshots of participants’ social behaviors in RemoteChess.}
    \Description{This figure was to explain how participants socialized with each other in RemoteChess. Screenshots of participants’ social behaviors in RemoteChess.}
    \label{fig:pin}
\end{figure*}

\subsubsection{\rv{Conversation Content}}

\rv{We focused on the conversations among participants because they play a significant role in social interactions. After analysis, we found that participants' conversation content varied across the two scenes with different social tendencies. We summarized several main conversational themes among the participants, including \textbf{chess-related} (Scene 2), \textbf{culture-related} (Scene 1), \textbf{life-related} (Scene 1), and \textbf{VR-related} (Scene 1, Scene 2).
In Scene 1, participants conducted conversations around the exhibited cultural content. The scope of the conversations was extended to topics related to memories, culture, and art. For instance, P9 shared her personal story triggered by the cultural exhibition with other participants (P7, P8): \textit{"Seeing the painting reminded me of my childhood. My family was poor at that time. Our grandfather carved the chess board by himself and taught us to play Chinese chess."} P2 said: \textit{"I never knew Chinese chess had such a rich history,"} to which P1 responded: \textit{"Yes, there’s so much about it. I know a Chinese chess museum, maybe we could go there together this weekend."}
In Scene 2, participants' conversations were primarily focused on the game itself, such as strategies and advice. They discussed tactics, moves, and current situation. P6, as a chess spectator, said: \textit{"That move was a mistake. You’re setting yourself up for a big loss."} P18, while observing the game, commented: \textit{"You should have moved your horse to the left, that’s a better defense."}
Across both scenes, participants discussed their feelings about this VR experience. P10 remarked: \textit{"The visuals are so vivid. It feels like being in a real park."} P12 responded: \textit{"Yes, you can see the wind blowing the leaves off the trees and the water.”}}

\subsubsection{Participants' Social Behavior in RemoteChess}

In RemoteChess, participants engaged not only in physical interactions with objects (e.g., chessboard, chess pieces, and exhibits), but also in meaningful social interactions with other participants. Their behaviors varied between Scene 1 and Scene 2.
In Scene 1, participants' interactions were more diverse and dynamic. They discussed the cultural exhibits with each other as they explored the virtual space together (Figure~\ref{fig:pin}(f)). They followed each other while touring the exhibit (Figure~\ref{fig:pin}(g)), and upon encountering one another, they frequently greeted each other through gestures such as waving (Figure~\ref{fig:pin}(d)), patting a virtual shoulder, or attempting a virtual handshake (Figure~\ref{fig:pin}(e)). These non-verbal actions reflected the preservation of real-world social etiquette and habits in the virtual reality environment, which enhanced their sense of immersion and social engagement.
In Scene 2, participants in the spectator role commented on the chess moves (Figure~\ref{fig:pin}(c)), while the players chatted as they played Chinese chess (Figure~\ref{fig:pin}(b)). Additionally, participants casually strolled through the virtual park with their companions (Figure~\ref{fig:pin}(a)). Although verbal communication was the primary form of interaction, we also observed various non-verbal behaviors. These non-verbal behaviors indicate that RemoteChess encourages a range of interaction methods, fostering social connections in VR not only through verbal communication but also through gestures and physical actions, enriching the overall social experience.

\subsubsection{Asymmetric character design inspires diverse interaction styles}

RemoteChess introduced asymmetric roles in Scene 2 (i.e., chess player and chess spectator). This theme presented participants' positive attitudes toward this setup, as well as different preferences for the two roles, with some participants desiring more interactive features.
Most participants (N=13) felt that the addition of chess spectator created a more authentic Chinese chess atmosphere. 
P6 mentioned: \textit{"Compared to online chess apps on my phone, I prefer this experience where I can both play and watch chess. It feels more like in real life."} Participants appreciated the chess spectator role as a new way to engage within the game. P8 said: \textit{"I enjoy being a spectator in the Chinese chess game. It feels like I’m an ‘invisible player,’ offering advice to those playing."} 
Similarly, P6 noted: \textit{"By giving advice and comments, I can influence the flow of the game." } These insights suggest that spectators are not just passive observers but active participants who can engage through discussion and suggestions, enhancing the interactivity of the game.

Participants showed varied preferences for the roles of a chess player or chess spectator. For example, P7, P9, and P15 preferred being chess spectators. P9 explained: \textit{"Watching the game helps me understand people’s personalities. I can also learn strategies that will help me improve my own game."} On the other hand, some participants preferred playing the game themselves. P16 stated: \textit{"Playing Chinese chess engages my mind more, and I feel more involved."} This sentiment was echoed by P10 and P11. P10 remarked: \textit{"Playing is much more exciting than just watching."} These differences reflect the different experiences participants sought in the game: some preferred the direct challenge of thinking and making decisions, while others preferred to observe and analyze. Additionally, some participants expressed a desire for more interactive features for the chess spectator role. For instance, P12 suggested adding virtual expressions or other interactive tools to make the spectator’s participation in discussions and commentary more dynamic. This feedback indicates that future designs could further enhance the spectator’s engagement and enrich the overall game experience by expanding the interactive options.

\section{Discussion}

Our work proposed RemoteChess, which constructs a social community and a culturally relevant activity (i.e., Chinese Chess) to enhance social connectedness among older adults. In this section, we discuss the key contributions embodied in RemoteChess design and how these findings can contribute to future social VR design. Also, we summarize design guidelines for designing culturally relevant social activities in VR to promote elder connectedness in other contexts. 

\subsection{Constructing VR Community for Older Adults}

Previous research in the HCI field primarily focused on promoting people’s social connections through remote gaming \cite{yuan2021tabletop,mills2023remote,sykownik2023vr}. Different from prior work, we construct a VR community that enables older adults to explore a more open virtual environment while participating in games. The community consists of two scenes \rv{with different social tendencies}: Scene 2 lets participants play Chinese chess games, while Scene 1 provides them with a space for free exploration and social interaction. This setting received positive feedback from our participants. They felt that combining games with an open environment not only provides a venue for social interaction, but also encourages users to freely explore the virtual environment, increasing the diversity of social interaction \sout{and cognitive experience}. This finding aligns with Xu et al.'s work, which suggests that open or sandbox environments, such as Second Life and Minecraft, are more effective in fostering social interactions in VR \cite{xu2023designing}. Because they allow users to explore, interact, and discover together, rather than passively accepting a predetermined storyline or experience \cite{xu2023designing}. Future work could further explore how these open-ended interaction scenarios can be extended in VR to provide more flexible and rich social experiences for older adults. \rv{Additionally, participants' conversation content aligns with the social tendencies of the two scenes. Scene 1, as a topic-oriented social scene, encouraged broader discussions, often extending beyond the immediate context to include personal stories and shared memories. In contrast, Scene 2, as an activity-oriented social scene, fostered more focused dialogues centered on the gameplay itself. Future work could consider integrating diverse social tendencies into VR, thus offering both structured and unstructured interaction opportunities to cater people's varying social preferences.}

Community-based social programs play a vital role in older adults' daily lives by providing them with rich social opportunities \cite{agedcareguide_social_support,lindsay2018mixed,xing2023keeping}. Based on a real-world social program, our work builds a VR social community for older adults. Our findings show that participants displayed high engagement when experiencing familiar social activities with others in the VR community. This not only helped them maintain social connections but also made it easier for them to adapt to the new VR interfaces. This aligns with previous research indicating that older adults are interested in extending their physical lives into virtual worlds through activities resembling real-life experiences \cite{siriaraya2012characteristics}. Additionally, our study revealed several other social programs participants wanted to engage in, such as cooking, opera, and crafting. Future work could explore how to immersively implement these activities in VR and whether they further enhance social connections and interaction among older adults.

\subsection{Design Guidelines for Creating Culturally Relevant Activities in VR to Foster Elder Connectedness}

While previous work has explored the potential of VR to support social connectedness among older adults \cite{baker2021school,baker2019interrogating,afifi2023using,wei2023bridging}, to our knowledge, we were the first to use a culturally relevant activity to construct a VR system for elder connectedness. Our findings demonstrate the value of creating culturally relevant activities in VR to enhance older adults' social connectedness. However, our work focuses on the specific case of Chinese chess in the Chinese context. On a global scale, the cultural \sout{and social} contexts of other countries must also be considered. Cultural activities may vary across regions, but their core value lies in \sout{enhancing social interaction and} fostering a sense of belonging through shared and familiar cultural elements \cite{huang2008social}. Based on our results, we propose the following design guidelines to inform future work in creating culturally relevant activities in VR to foster elder connectedness. \rv{We also discussed how these guidelines could be adapted for different cultural contexts.}

\subsubsection{\rv{Design Guideline 1:} Resemble familiar settings related to the social activity in virtual environments}

RemoteChess uses a virtual \sout{outdoor} park to replicate a familiar environment where older adults gather to play Chinese chess, along with background sounds of others playing and natural elements like wind through trees. \sout{Our findings show that familiar settings related to the activity promoted users' social interaction in VR. Participants reported that the park scene made them feel more immersed and more willing to engage with others.} \rv{Our participants reported that the familiar settings related to the activity made them feel more immersed and more willing to engage with others.} This aligns with prior work which suggests familiar real-world settings can \sout{influence the quality of} \rv{promote users'} social interactions in VR \cite{mcveigh2019shaping}. \sout{Despite these benefits, participants expressed a desire for more multi-sensory elements. They envisioned smelling flowers in the park (olfactory), feeling the texture of chess pieces (tactile), and sensing warmth when shaking hands with others (tactile). Recent studies suggest that multi-sensory experiences in VR can enhance immersion, improve cognitive function, and increase users' acceptance and engagement with VR through activities like games or reminiscence therapy, especially for older adults \cite{li2024designing, Lee2023}.} \rv{However, participants expressed a desire for more multi-sensory elements, such as the ability to smell the flowers or feel the texture of chess pieces, which could enhance users' immersion and engagement with VR \cite{li2024designing, Lee2023}.} Future work could prioritize resembling familiar social scenes closely tied to cultural activities in VR, integrating multi-sensory elements to further enhance people's familiarity and immersion.
\rv{Under other cultural contexts, VR design can better reflect the local familiar environments associated with specific activities. For instance, in India, a virtual courtyard where older adults gather to play Carrom could be used, while in Spain, a virtual plaza could host activities like Domino games. These environments would evoke familiar cultural memories, enhancing immersion across different cultural contexts.}

\subsubsection{\rv{Design Guideline 2:} Provide multiple levels of engagement to meet different social needs}

Social presence in VR is influenced by symmetrical and asymmetrical interactions \cite{yassien2020design,drey2022towards,dubosc2021impact}. To cater the need of many older adults who like watching Chinese chess, RemoteChess introduced asymmetrical roles (chess player and chess spectator) in Scene 2. Our results demonstrated the benefits of this approach,  which allows participants with different personalities and preferences to find their own comfortable way to engage in social interaction. This aligns with Harris et al.'s findings that asymmetry can be used as a design tool to promote social interaction between players \cite{harris2019asymmetry}. Future work could expand this interaction model by introducing more roles and modes of interaction, such as performer and audience, to offer diverse participation options and meet the needs of different older adults. \sout{Participants also suggested adding more interactive options for non-players (P12) to enhance their sense of involvement. Future work could further enhance the overall experience of watching asymmetric systems by enriching the interaction capacities of non-players.}
\rv{This guideline can also be extended to other cultural contexts by adjusting the roles and participation modes according to local activities. In many Western cultures, Bridge is a highly popular social game among older adults. In VR, a community focused on card games could be simulated, allowing older adults with mobility or distance constraints to easily participate in both games and social gatherings. By enabling older adults to engage either as players or spectators, VR can cater to the diverse needs of older adults, whether they prefer active participation or more passive roles.}

\subsubsection{Design Guideline 3: Incorporate social catalysts to inspire the conversation}
Social catalysts, such as focus objects, can prompt conversations and interactions between remote users \cite{mcveigh2019shaping,karahalios2004telemurals,roth2018beyond}. In RemoteChess, a cultural exhibition related to Chinese chess served as a social catalyst, sparking discussions among participants. Our findings show that these social catalysts enhanced social interactions in several ways: 1) provided engaging topics for discussion, 2) fostered a sense of community, 3) evoked shared memories, and 4) encouraged the sharing of personal stories. Future work could use similar social catalysts under other cultural contexts, such as themed exhibits or displays related to the activity. Participants also suggested more dynamic social catalysts, such as cultural performances in VR where they could take on different roles. They thought this would not only inspire discussion but also deepen interactions through shared experiences. Future work could investigate how to present more interactive and engaging social catalysts in VR.

\section{Limitation \& Future Work}

One limitation was that our work only assessed the immediate sense of social connection felt by older adults after short-term use of the system. While participants showed a positive attitude toward RemoteChess's capability, they only experienced the system for 45 minutes during the study. Social connections are usually established through long-term interactions. Future work should consider conducting long-term studies to observe the sustained impact of such VR systems on older adults' social connectedness.
\rv{Moreover, familiarity among participants is also an important factor influencing social interactions. Future research could investigate how varying levels of familiarity among participants, ranging from strangers to close acquaintances, affect the quality and depth of interactions within the VR community.}

Second, as our participants are all from a single cultural background, the generalizability of the results may be limited by cultural differences. Future research should extend this work to older adults from other cultural backgrounds and explore how different cultural activities may facilitate social interactions in VR.

Finally, although RemoteChess successfully promoted social interaction among older adults through Chinese chess, its single cultural activity format may limit users' long-term interest and participation. Future work could explore a wider variety of cultural activities, such as music, calligraphy, or other highly interactive cultural projects. This would help cater to the interests of older adults from different backgrounds and provide a richer social experience.

\section{Conclusion}

In this paper, we explored the potential of using culturally relevant activities in VR to enhance social connectedness among older adults. We implemented RemoteChess, a VR social community that promotes social interaction through familiar environments, cultural elements, and asymmetric game roles. In an evaluation with 18 older adults, we found that familiar settings related to the social activity successfully stimulated social engagement, while the asymmetric player-spectator roles fostered a dynamic social environment. Furthermore, using cultural elements as conversation starters contributed to enhanced feelings of social presence and community belonging among participants. Based on our findings, we contribute design guidelines for future social VR design aimed at social connectedness of older adults. Overall, our study demonstrates the potential of culturally relevant activities in social VR as a promising avenue for enhancing older adults’ well-being through remote social engagement.

\begin{acks}
This work is partially supported by the Guangzhou-HKUST(GZ) Joint Funding Project (No. 2024A03J0617),  Education Bureau of Guangzhou Municipality Funding Project (No. 2024312152), Guangzhou Higher Education Teaching Quality and Teaching Reform Project (No. 2024YBJG070), Guangdong Provincial Key Lab of Integrated Communication, Sensing and Computation for Ubiquitous Internet of Things (No. 2023B1212010007), the Project of DEGP (No.2023KCXTD042), and the Guangzhou Science and Technology Program City-University Joint Funding Project (No. 2023A03J0001).
\end{acks}

\bibliographystyle{ACM-Reference-Format}
\bibliography{main}

\end{document}